\documentclass[reprint,aip,jcp]{revtex4-2}
\usepackage{amsfonts}
\usepackage{amsmath}
\usepackage{amssymb}
\usepackage{graphicx}
\usepackage{dcolumn}
\usepackage{bm}
\usepackage{float}

\begin{document}

\title{Capturing anharmonic effects in single vibronic level fluorescence spectra
using local harmonic Hagedorn wavepacket dynamics}

\author{Zhan Tong Zhang}
\affiliation{Laboratory of Theoretical Physical Chemistry, Institut des Sciences et
Ing\'enierie Chimiques, Ecole Polytechnique F\'ed\'erale de Lausanne (EPFL),
CH-1015 Lausanne, Switzerland}
\author{M\'at\'e Visegr\'adi}
\affiliation{Laboratory of Theoretical Physical Chemistry, Institut des Sciences et
Ing\'enierie Chimiques, Ecole Polytechnique F\'ed\'erale de Lausanne (EPFL),
CH-1015 Lausanne, Switzerland}
\author{Ji\v{r}\'i J. L. Van\'i\v{c}ek}
\email{jiri.vanicek@epfl.ch}
\affiliation{Laboratory of Theoretical Physical Chemistry, Institut des Sciences et
Ing\'enierie Chimiques, Ecole Polytechnique F\'ed\'erale de Lausanne (EPFL),
CH-1015 Lausanne, Switzerland}
\date{\today}

\begin{abstract}

Hagedorn wavepacket dynamics yields exact single vibronic level (SVL) fluorescence spectra in global harmonic models.
To partially describe the effects of anharmonicity, important in the spectra of real molecules, we describe a combination of the Hagedorn wavepacket approach to SVL spectroscopy with the local harmonic approximation. In a proof-of-principle study [Phys. Rev.~A~\textbf{111}, L010801 (2025)], we
successfully demonstrated the utility of this method by computing the SVL spectra of difluorocarbene, a floppy molecule with moderately anharmonic potential.
Here, we describe the theory in detail and analyse the method more thoroughly. To assess the accuracy of the method independently of electronic structure errors, we use a two-dimensional Morse-type potential for which exact quantum benchmarks are available, and show that the local harmonic approach yields more accurate results than global harmonic approximations, especially for the emission spectra from higher initial vibrational levels.  Next, we compare the global and local harmonic SVL spectra of anthracene, where the more expensive local harmonic corrections turn out to be less important as long as the correct global harmonic model is used. We also present additional local harmonic results for difluorocarbene, where treating anharmonicity is essential for accurate evaluation of the spectra. Yet, we also show that the structure of the difluorocarbene spectra can be explained qualitatively (but not quantitatively) with a reduced-dimensional harmonic model, for which the spectral intensities can be evaluated analytically.
\end{abstract}

\maketitle

\section{Introduction}

In the time-dependent approach to spectroscopy \cite{Heller:1981a}, vibrationally resolved electronic spectra can frequently be described by the dynamics of a Gaussian wavepacket \cite{Heller:1975,Wehrle_Vanicek:2014}. Under several reasonable assumptions (i.e., validity of the time-dependent perturbation theory and Born-Oppenheimer, low-temperature, and Condon approximations), the Gaussian wavepacket dynamics becomes exact if the involved potential energy surfaces are harmonic. In many situations, however, one needs to propagate \emph{non-Gaussian} wavepackets \emph{even if} the potentials are harmonic. The single vibronic level (SVL) fluorescence spectroscopy provides one of the simplest examples where this happens, and therefore we only discuss this technique even though the Hagedorn-wavepacket-based approach described below would be very useful for other spectroscopy techniques requiring the propagation of non-Gaussian wavepackets. 

The SVL fluorescence spectroscopy has been used to study intramolecular relaxation \cite{Parmenter_Schuyler:1970,Lambert_Zewail:1984a,Felker_Zewail:1985,Numata_Suzuka:2012}, characterise molecular vibronic structure \cite{Quack_Stockburger:1972,Suzuki_Okuyama:2024}, and identify  conformers and reactive intermediates \cite{Hollas_BinHussein:1989,Newby_Zwier:2009,Smith_Clouthier:2022a}.
Recently, several time-dependent approaches have been developed to simulate SVL spectra in global harmonic models \cite{Huh_Berger:2012,Tapavicza:2019,Zhang_Vanicek:2024a,Zhang_Vanicek:2025a}. Tapavicza \cite{Tapavicza:2019} used a generating function formalism to compute the SVL spectra of anthracene from singly excited levels using a harmonic model determined from accurate electronic structure calculations.
Motivated by Tapavicza's work, we proposed a method based on Hagedorn wavepacket dynamics \cite{Hagedorn:1998,Faou_Lubich:2009,Lasser_Lubich:2020},
which enabled us to compute the SVL spectra from any initial vibrational levels \cite{Zhang_Vanicek:2024a} and to simulate the SVL spectra of higher vibrational levels of anthracene within the global harmonic approximation \cite{Zhang_Vanicek:2025a}.

To accurately compute vibrationally resolved spectra of polyatomic molecules, however, it is often necessary to account for the anharmonicity of molecular potential energy surfaces (PESs) \cite{Chau_Mok:2001,Luis_Kirtman:2004,Bonness_Luis:2006,GalestianPour_Hauer:2017,Bonfanti_Pollak:2018,Barone_Puzzarini:2021,Conte_Ceotto:2023}. For flexible molecules with large-amplitude anharmonic vibrational motion or when highly accurate (ro)vibrational levels are needed, a quantum treatment is often required \cite{Bacic_Light:1986,Bacic_Light:1987,Bacic_Light:1988,Felker_Bacic:2019}. Conventional time-independent methods that evaluate Franck--Condon factors for each transition can include anharmonic effects by using the variational principle \cite{Bacic_Light:1986,Luis_Christiansen:2006,Bowman_Meyer:2008,Meier_Rauhut:2015,Costa_Vidal:2018} or perturbation theory \cite{Luis_Kirtman:2004,Krasnoshchekov_Stepanov:2015,Fuse_Bloino:2024}.
In some cases, the use of curvilinear coordinates can provide more accurate representations of molecular motions and improve the description of anharmonicity \cite{Bacic_Light:1989,Reimers:2001,Borrelli_Peluso:2006,Capobianco_Peluso:2012,Mendolicchio:2023}.
In higher-dimensional systems with Duschinsky coupling, the determination of anharmonic wavefunctions and the computation of a large number of Franck--Condon overlaps become very costly, if not unfeasible.
Even when such calculations are feasible, computing all Franck--Condon factors is wasteful when individual peaks are not visible in low- or intermediate-resolution spectra.

Alternatively, in the time-dependent framework, the spectrum is computed as the Fourier transform of an appropriate autocorrelation function obtained by propagating a wavepacket on the final electronic PES \cite{Heller:1981a,book_Tannor:2007}.
When simulating electronic spectra at vibrational resolution,
semiclassical trajectory-based methods can capture the effects of an anharmonic PES in larger systems more efficiently using an on-the-fly implementation that relies only on the local PES information.
In the thawed Gaussian approximation \cite{Heller:1975,Grossmann:2006,Wehrle_Vanicek:2014,Vanicek_Begusic:2021} (TGA), a Gaussian wavepacket is propagated using the local harmonic approximation (LHA), which expands the true potential locally to the second order at each time step. The centre of the wavepacket then follows the anharmonic classical trajectory, while the width evolves according to the local Hessian matrix.

Whereas the TGA uses a single Gaussian wavepacket to simulate the emission or absorption process from the ground vibrational level \cite{Wehrle_Vanicek:2014,Wehrle_Vanicek:2015,Begusic_Vanicek:2019,Kletnieks_Vanicek:2023}, our approach \cite{Zhang_Vanicek:2024a,Zhang_Vanicek:2025,Zhang_Vanicek:2025a}, based on Hagedorn wavepackets \cite{Hagedorn:1998,Faou_Lubich:2009,book_Lubich:2008,Lasser_Lubich:2020}, uses functions that are polynomials multiplied by a Gaussian to compute SVL spectra from any vibrationally excited levels. These functions result from the application of a special raising operator to a Gaussian wavepacket and are not simple products of Hermite functions in higher dimensions. The Hagedorn functions are exact solutions of the time-dependent Schr\"{o}dinger equation (TDSE) not only in global but also in local harmonic models \cite{Lasser_Lubich:2020}. Combination of Hagedorn wavepacket dynamics with the local harmonic approximation makes it possible to partially account for the anharmonic effects in SVL spectra.
Moreover, the local harmonic Hagedorn dynamics retains the ability to obtain SVL spectra from arbitrary vibrational levels from a single, common Gaussian trajectory, making our method especially appealing for combination with on-the-fly ab initio dynamics.

In a letter published in Ref.~\cite{Zhang_Vanicek:2025}, we demonstrated that the on-the-fly local harmonic Hagedorn approach is effective for evaluating SVL spectra of certain floppy molecules, such as difluorocarbene, at least at lower vibrational excitations. Here, we describe the theory and analyse the local harmonic Hagedorn wavepacket approach to spectroscopy in more detail. To validate the method independently of electronic structure errors, we compare in Sec.~\ref{subsec:2d} its results with exact quantum SVL spectra for a two-dimensional Morse-type system. The local harmonic spectra are substantially better than both the vertical and adiabatic global harmonic spectra; however, significant differences from the exact results still exist for the spectra with high initial excitations.
In Sec.~\ref{subsec:anthracene}, we analyse the effects of anharmonicity in anthracene, which was studied in Ref.~\cite{Zhang_Vanicek:2025a} within the global harmonic approximation. 
Finally, in Sec.~\ref{subsec:cf2}, we present additional results for the much smaller but floppier difluorocarbene, previously studied in Ref.~\cite{Zhang_Vanicek:2025}. On one hand, we confirm the necessity of going beyond the global harmonic approximation if one needs accurate positions and intensities of spectral peaks. On the other hand, we qualitatively explain the origin of the spectral envelope splitting induced by vibrational excitation in difluorocarbene using a one-dimensional harmonic model.

\section{Theory}
\label{sec:theory}

In single vibronic level fluorescence spectroscopy, the molecular population in the ground electronic state $g$ is first excited to a specific vibrational level $|K\rangle \equiv |e, K\rangle$ in an electronically excited state $e$, where $K=(K_1,\dots,K_D)$ is a multi-index of non-negative integers specifying the vibrational quantum numbers in the $D$ vibrational degrees of freedom. The rate of subsequent spontaneous emission from this vibronic level with energy $\hbar\omega_{e,K}$ is given by the Fourier transform \cite{Heller:1981a,book_Tannor:2007,Tapavicza:2019} 
\begin{equation}
    \sigma_{\text{em}}(\omega) = \frac{4\omega^3}{3\pi \hbar c^3} |{\mu}%
    _{ge}|^2 \operatorname{Re} \int^\infty_{0} \overline{C(t)} \exp[it(\omega -
   \omega_{e,K})] \,dt
\end{equation}
of the wavepacket autocorrelation function
\begin{equation}
    C(t)= \langle K | \exp(-i H_{g}t/\hbar) | K \rangle,
    \label{eq:autocorr}
\end{equation}
obtained by propagating the initial wavepacket with the ground-state Hamiltonian $H_{g}$. We use the Condon approximation, in which the transition dipole moment $\mu_{ge}$ does not depend on the nuclear coordinates. 
Here, we consider only the SVL emission without regard to competing intramolecular relaxation and energy redistribution processes in the excited electronic state.

Assuming the harmonic approximation for the excited-state surface $V_{e}$, we can represent the initial vibrational wavepacket $|K\rangle$ by a Hagedorn function \cite{Zhang_Vanicek:2024a}
\begin{equation}
    \varphi_{K} = ({K!})^{-1/2} (A^{\dagger})^{K} \varphi_0, \label{eq:hgf}
\end{equation}
obtained by applying Hagedorn's raising operator \cite{Hagedorn:1998}
\begin{equation}
    A^{\dagger}  := \frac{i}{\sqrt{2 \hbar}}
    \left(P_{t}^{\dagger} \cdot (\hat{q} - q_{t})  - Q_{t}^{\dagger} \cdot (\hat{p} - p_t) \right)
    \label{eq:A_dag}
\end{equation}
to the $D$-dimensional normalised Gaussian parameterised in the excited-state normal-mode coordinates as
\begin{multline}
    \varphi_{0}(q) = \frac{1}{(\pi \hbar)^{D / 4} \sqrt{\det (Q_{t})%
    }} \\
    \times \exp \left\{ \frac{i}{\hbar} \left[ \frac{1}{2} x^{T} \cdot P_{t}
    \cdot Q_{t}^{- 1} \cdot x + p_{t}^{T} \cdot x + S_{t} \right] \right\}.  \label{eq:tga}
\end{multline}
Here, $x:= q - q_{t}$ is the shifted position, $q_{t}$ and $p_{t}$ are the position and momentum of the
centre of the wavepacket in the phase space, and $S_{t}$ is the classical action. Compared with Heller's parametrisation \cite{Heller:1975,Vanicek:2023}, the width matrix $\mathrm{A}_{t} \equiv P_{t}\cdot Q_{t}^{-1}$ of the Gaussian is factorised in terms of two complex-valued $D$-dimensional matrices $Q_{t}$ and $P_{t}$, which satisfy the symplecticity
conditions~\cite{book_Lubich:2008,Lasser_Lubich:2020}
\begin{align}
Q_{t}^{T} \cdot P_{t} - P_{t}^{T} \cdot Q_{t} &= 0 \qquad{\text{and}}  \label{eqn:symp_rel1} \\
Q_{t}^{\dagger} \cdot P_{t} - P_{t}^{\dagger} \cdot Q_{t} & = 2 i \mathrm{Id},
\label{eqn:symp_rel2}
\end{align}
where $\mathrm{Id}$ is the $D$-dimensional identity matrix. 
This factorisation allows the construction of a multi-dimensional raising operator, and yields simpler equations of motion that are linear in $Q$ and $P$, analogous to the linearized classical equations for position and momentum \cite{Hagedorn:1998,Faou_Lubich:2009} [see Equations (\ref{eq:eom}) below]. Moreover, the symplectic structure of Gaussian wavepackets becomes canonical with Hagedorn’s parametrisation \cite{Ohsawa_Leok:2013,Ohsawa:2019}.
In Equation~(\ref{eq:hgf}), we used the multi-index notation for $K! = K_{1}!\cdot K_{2}! \cdot \,\cdots\, \cdot K_{D}!$ and $(A^{\dagger})^{K}=(A^{\dagger}_{1})^{K_{1}}\cdot (A^{\dagger}_{2})^{K_{2}}\cdot\, \cdots\, \cdot (A^{\dagger}_{D})^{K_{D}}$, where $A^{\dagger}_{j}$ is the $j$-th component of the vector operator $A^{\dagger}$.

In position representation, the Hagedorn functions~(\ref{eq:hgf}) are given by polynomials multiplied by a common Gaussian~(\ref{eq:tga}) and form a complete orthonormal basis in $L^2(\mathbb{R}^D)$. They can therefore be used to approximate an arbitrary wavepacket solution to the TDSE with the linear combination
\begin{equation}
    \Psi = e^{iS_t/\hbar}\sum_{K\in\mathcal{K}} c_K \varphi_K,\label{eq:hwp}
\end{equation}
where $c_K$ is a complex coefficient and $\mathcal{K}$ is a truncated set of multi-indices in $\mathbb{N}_{0}^D$.
However, instead of using Hagedorn functions as a time-dependent basis as in previous applications \cite{Hagedorn_Joye:2000,Bourquin_Hagedorn:2012,Kieri_Karlsson:2012,Zhou:2014,Gradinaru_Rietmann:2024}, here we exploit the fact that they are exact solutions to the TDSE with not only a global harmonic but also a local harmonic potential, which makes it possible to partially capture the anharmonic effects. 

In the LHA, the true potential $V$ is replaced with an effective quadratic potential
\begin{equation}
    V_{\text{LHA}}(q; q_t) := V(q_{t}) + V^{\prime}(q_{t})\cdot x + x^{T}\cdot V^{\prime\prime}(q_{t}) \cdot x/2,
\end{equation}
given by the second-order Taylor expansion of $V$ around the centre $q_{t}$ of the wavepacket at each time. In this effective potential, an initial Gaussian wavepacket (\ref{eq:tga}) retains its Gaussian form at all times, and its parameters evolve according to the equations of motion
\begin{align}
\dot{q_t} &= m^{-1} \cdot p_t, & \dot{p_t} &= -V^{\prime}(q_t)\nonumber \\ 
\dot{Q_t} &= m^{-1} \cdot P_t, & \dot{P_t} &= -V^{\prime\prime}(q_t) \cdot Q_t,\nonumber \\
\dot{S_t} &= L_t,\label{eq:eom}
\end{align}
where $m$ is the real symmetric mass matrix and $L_{t} = p^{T} \cdot m^{-1} \cdot p / 2-V_g(q_{t})$ is the Lagrangian \cite{book_Lubich:2008,Lasser_Lubich:2020}.

Due to Hagedorn's construction of the ladder operators, evolving the full Hagedorn wavepacket~(\ref{eq:hwp}) also becomes particularly simple with $V_{\text{LHA}}$. Notably, the coefficients $c_K$ in the wavepacket (\ref{eq:hwp}) do not change at all within the local harmonic approximation.  
 Specifically for SVL processes, the propagated state remains represented by a single Hagedorn function at all times.  
 Although the index $K$, representing the initial vibrational excitation in the excited electronic state, remains fixed during the local harmonic dynamics on the final (ground-state) surface, the evolution of the wavepacket encodes in the time domain transitions to all accessible vibrational levels of the ground electronic state \cite{Heller:1981a}.

To show that the coefficients $c_{K}$ do not evolve under the LHA, one notes
that Hagedorn's raising operator (\ref{eq:A_dag}) satisfies the equation of motion \cite{Lasser_Lubich:2020}
\begin{equation}
i\hbar\dot{A}_{t}^{\dagger}=-[A_{t}^{\dagger},\hat{H}_{\text{LHA}}(\varphi
_{0})]\label{eq:Adag_dot}
\end{equation}
with a local harmonic Hamiltonian $\hat{H}_{\text{LHA}}(\varphi_{0})=\hat
{p}\cdot m^{-1}\cdot\hat{p}/2+V_{\text{LHA}}(\hat{q};q_{t})$, which can be
proven by evaluating both sides and using the equations of motion (\ref{eq:eom}) for the
parameters of the guiding Gaussian $\varphi_{0}$. Then one demonstrates that
Hagedorn functions are exact solutions of a nonlinear TDSE
\begin{equation}
i\hbar\dot{\varphi}_{K}=\hat{H}_{\text{LHA}}(\varphi_{0})\varphi
_{K}.\label{eq:TDSE_HF_Heff}
\end{equation}
For the ground Hagedorn function $\varphi_{0}$ described by a Gaussian wavepacket, this is well known as the
thawed Gaussian approximation \cite{Heller:1975}. For excited
Hagedorn functions $\varphi_{K}$, it can be proven by induction on multi-index $K$ using the equation of
motion (\ref{eq:Adag_dot}). Because Hagedorn functions solve the TDSE with $\hat{H}_{\text{LHA}}$ independently, the Hagedorn wavepacket~(\ref{eq:hwp}) satisfies the same equation if and only if the coefficients $c_{K}$ are time-independent.

Consequently, the propagation of any SVL initial state (\ref{eq:hgf}) does not require solving any additional equations of motion beyond Equations~(\ref{eq:eom}) for a Gaussian wavepacket and does not depend on the initial vibrational excitation $K$. The SVL spectra from all vibrational levels can thus be obtained from a single Gaussian trajectory, and in \textit{ab initio} applications, expensive on-the-fly electronic structure calculations do not have to be repeated for different $K$.

Equations~(\ref{eq:eom}) are deceptively similar to those for the Hagedorn dynamics in global harmonic models \cite{Zhang_Vanicek:2024a}, the only difference being that Equations~(\ref{eq:eom}) contain the true potential $V$, its gradient $V^{\prime}$, and Hessian $V^{\prime\prime}$ instead of the corresponding derivatives of the global harmonic approximation $V_{\text{HA}}$. As a result, the local harmonic approximation does capture some anharmonicity. However, whereas the global harmonic case can be solved analytically for arbitrary times \cite{book_Tannor:2007}, the local harmonic dynamics necessitates numerical propagation. In this approach, the centre of the wavepacket is guided by the classical trajectory and can fully explore the anharmonic shape of the PES.

Our approach fully incorporates Duschinsky rotation effects and, by a straightforward extension of the global harmonic equations, partially includes anharmonicity. This simplicity underlies the elegance in the construction of the Hagedorn functions. Compared to other methods based on Hermite or other `polynomial-times-Gaussian' bases \cite{Billing:1999,Borrelli_Peluso:2006,Faou_Gradinaru:2008}, which may provide greater accuracy but are more complex due to the need for propagating the coefficients, our local harmonic Hagedorn method offers a more accessible and efficient alternative.

To evaluate the autocorrelation function (\ref{eq:autocorr}), it is necessary to compute the overlaps of the initial and final Hagedorn functions.
The explicit form of the polynomial factor in a multidimensional Hagedorn function is not known, and we did not find a simple, closed-form expression for the overlaps of Hagedorn functions associated with different Gaussians.
Instead, we used the exact recursive algebraic expressions for these overlaps derived in Ref.~\cite{Vanicek_Zhang:2025}, thereby avoiding errors and difficulties arising from the use of numerical quadratures \cite{Bourquin:2017}.

The ab initio application of local harmonic Hagedorn wavepacket dynamics to molecular systems requires on-the-fly evaluations of the potential energy as well as its gradients and Hessians. When the global harmonic approximation is used, the results from electronic structure programs, typically performed in Cartesian or internal coordinates, are converted to normal-mode coordinates to construct the local harmonic potential for the subsequent vibrational nuclear dynamics. In the on-the-fly case, however, the propagated position of the wavepacket must be converted from normal-mode to Cartesian coordinates for the ab initio calculation at each time step, and the computed energy, gradient, and Hessian are then transformed back to normal-mode coordinates for wavepacket propagation (see section 6.7 of Ref.~\cite{Vanicek_Begusic:2021}).

\section{Results}

\subsection{Two-dimensional coupled Morse potential}
\label{subsec:2d}

\begin{figure*}
    \centering
    \includegraphics[width=\linewidth]{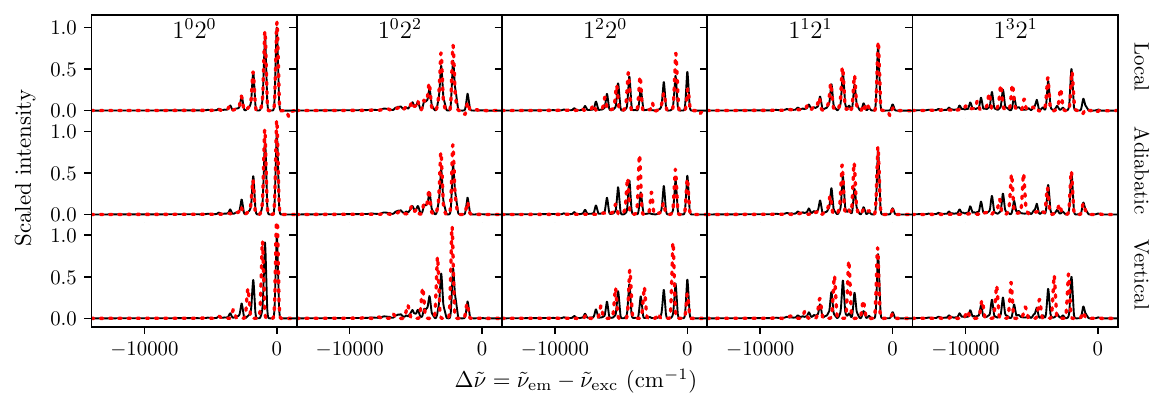}
    \caption{SVL spectra computed with Hagedorn wavepackets (red dashed lines) propagated using the local, adiabatic, or vertical harmonic approximation of a two-dimensional coupled Morse potential
    are compared to exact quantum spectra (black solid lines); the initial vibrational excitation is indicated at the top.}
    \label{fig:2d}
\end{figure*}

\begin{figure*}
    \centering
    \includegraphics[width=\linewidth]{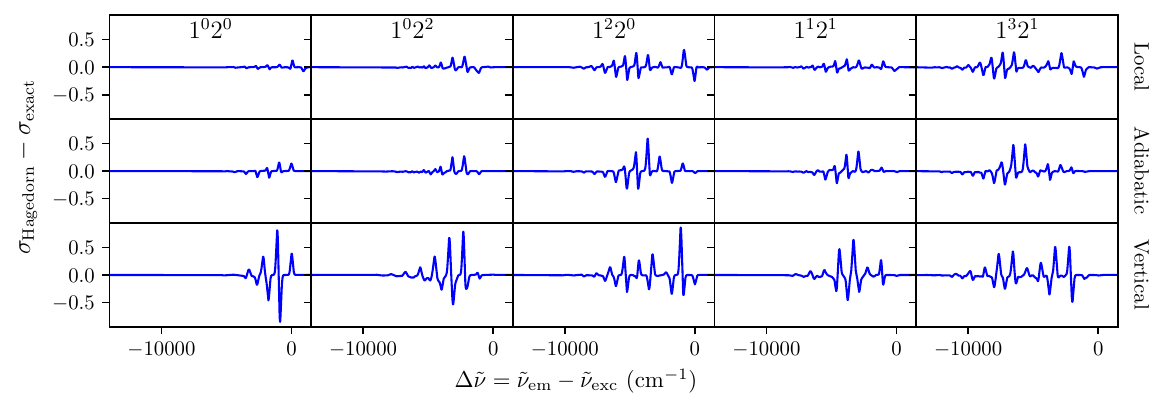}
    \caption{Differences ($\sigma_{\text{Hagedorn}} - \sigma_{\text{exact}}$) between the SVL spectra of a two-dimensional coupled Morse potential computed with Hagedorn wavepacket dynamics and the corresponding exact spectra. See the caption of Figure~\ref{fig:2d} for more details.}
    \label{fig:2da}
\end{figure*}

In Ref.~\cite{Zhang_Vanicek:2025}, we compared ab initio local harmonic Hagedorn calculations of difluorocarbene SVL spectra with experiment; however, the quantum chemistry calculations were a significant source of error in the observed discrepancies. Here, to study the local harmonic Hagedorn dynamics in a multidimensional anharmonic system without interference from electronic-structure (or experimental) errors, we performed numerical calculations in a two-dimensional model potential.

The Morse oscillator provides a simple, yet realistic one-dimensional anharmonic potential, 
\begin{equation}
V(q)=\frac{\omega_{g}}{4\chi}\left[ 1-e^{-\sqrt
{2m\omega_{g}\chi}(q-q_{\text{eq},g})}\right]^{2},\label{eq:morse_1d}
\end{equation}
where $q_{\text{eq},g}$ is the equilibrium position, $\omega_{g}$ is the frequency of the harmonic oscillator fitted at $q_{\text{eq},g}$, and $\chi$ is a dimensionless anharmonicity parameter.
However, in one dimension, Hagedorn functions are equivalent to the well-known Hermite functions, i.e., Hermite polynomials multiplied by a Gaussian \cite{Hagedorn:1998}. The beauty of Hagedorn functions emerges in higher dimensions, where they are not simple products of one-dimensional Hermite functions \cite{Lasser_Lubich:2020,Ohsawa:2019}. They remain exact solutions to the TDSE for both global and local harmonic potentials, even in the presence of Duschinsky rotation (mode mixing).

The coupled Morse potential provides a nonseparable generalization of the Morse potential to higher dimensions \cite{Fereidani_Vanicek:2023}. We assume that the final, ground electronic surface $V_{g}$ is given by
\begin{equation}
    V_{g}(q) = \sum_{j=1}^{D}V_{j}(q_j) + V_{\text{cpl}}(q),\label{eq:morse_nd}
\end{equation}
where for each vibrational degree of freedom $j$ a Morse term $V_{j}$ exists in the form of Equation~(\ref{eq:morse_1d}) specified by parameters $\omega_{g,j}$, $\chi_j$, and $q_{\text{eq},g,j}$. A $D$-dimensional coupling term,
\begin{equation}
    V_{\text{cpl}}(q) = d^{\prime} \left(
    1-e^{-a^{T} \cdot (q - q_{\text{eq},g})}\right)^{2},
\end{equation}
introduces nonseparability and depends on the dissociation energy $d^{\prime}$ and a vector $a \equiv (a_{1},\cdots,a_{D})$ of decay parameters, which are related by a dimensionless anharmonicity vector $\chi^{\prime} \equiv (\chi^{\prime}_{1},\cdots,\chi^{\prime}_{D})$ via
$
    a = \sqrt{8d^{\prime}}\, \chi^{\prime}
$.

A two-dimensional Morse system with $\omega_{g} = (0.0041, 0.005), \chi = (0.005, 0.002), q_{\text{eq},g} = (20, 5), d^{\prime} = 0.08$ and $\chi^{\prime} = (0.001,0.001)$ was chosen as a simple multidimensional example. For convenience, we set $\hbar=1$ in this model system and rescaled the coordinates so that each vibrational mode has the same effective mass $m=1$.
 The initial wavepacket $|K\rangle$ is assumed to be an eigenfunction of harmonic potential $V_{e}$ that is centered at $q_{\text{eq},e} = (0,0)$ and whose Hessian corresponds to fundamental frequencies $\omega_{e} = (0.00456, 0.00365)$.

To quantify anharmonic effects, two global harmonic models were constructed for $V_{g}$. In the vertical harmonic approximation (VHA), the effective potential was constructed around the Franck--Condon point as
\begin{multline}
    V_{\text{VHA}} = V_{g}(q_{\text{eq},e}) 
    + V^{\prime}_{g}(q_{\text{eq},e}) \cdot (q-q_{\text{eq},e})\\
    + (q-q_{\text{eq},e})^T\cdot V^{\prime\prime}_{g}(q_{\text{eq},e}) \cdot (q-q_{\text{eq},e})/2.\label{eq:vha}
\end{multline}
In the adiabatic harmonic approximation (AHA), the potential was expanded around the ground-state equilibrium position $q_{\text{eq},g}$ as
\begin{multline}
    V_{\text{AHA}} = V_{g}(q_{\text{eq},g})  \\
    +(q-q_{\text{eq},g})^T\cdot V^{\prime\prime}_{g}(q_{\text{ref},g}) \cdot (q-q_{\text{eq},g})/2, \label{eq:aha}
\end{multline}
where the gradient $V^{\prime}_{g}(q_{\text{eq},g})$ vanishes in the equilibrium configuration.

We used the local harmonic Hagedorn dynamics and the adiabatic and vertical harmonic approximations to compute the emission spectra from the ground level $1^{0}2^{0}$, doubly excited levels $1^{0}2^{2}$ and $1^{2}2^{0}$, and combination levels $1^{1}2^{1}$, and $1^{3}2^{1}$ (where the superscript in the mode label indicates the initial vibrational excitation in each mode).
Here we omit showing results for singly excited levels $1^{0}2^{1}$ and $1^{1}2^{0}$, which can also be obtained by other time-dependent approaches \cite{VonCosel_Burghardt:2017,Tapavicza:2019} in the harmonic approximation, and instead highlight the general applicability of our approach in the more challenging cases of multiply excited and combination levels.
The Hagedorn spectra are compared with the exact quantum results to assess the accuracy of the different (global or local harmonic) approximations to the anharmonic Morse potential. Split-operator quantum calculations were performed on a position grid with $256\times 256$ points ranging from $-128$ to $356$ in each dimension. 

In all simulations, the propagation lasted 80000\,a.u. (20000 steps with a time step of 4\,a.u.) and the correlation function (\ref{eq:autocorr}) was computed every five steps. Whereas the quantum calculation required a separate propagation of the wavepacket for each excitation level, in the Hagedorn approach it was sufficient to propagate only the ground-level Gaussian wavepacket to generate spectra for all excitation levels. The spectra were broadened by a Gaussian function with a half-width at half-maximum of $100\,\text{cm}^{-1}$. The intensities in the spectra were scaled by the intensity of the highest peak in the exact $1^0 2^0$ spectrum. 
The spectra are shown with respect to the difference $\Delta\tilde{\nu} = \tilde{\nu}_{\text{em}}  - \tilde{\nu}_{\text{exc}}$ between the wavenumbers $\tilde{\nu}_{\text{em}}$ of emission and $\tilde{\nu}_{\text{exc}}$ of the initial excitation; the transition to the ground level in the ground state is always at $0\,\text{cm}^{-1}$.

Figure~\ref{fig:2d} compares the spectra computed using different approximations with the exact quantum spectra, and the differences are shown in Figure~\ref{fig:2da}.
For the ground-level ($1^0 2^0$) emission spectra (first column in Figs.~\ref{fig:2d} and \ref{fig:2da}), both the adiabatic and local harmonic models show excellent agreement with the quantum results. The vertical model also performs reasonably well; however, the spacing between the peaks differs from the exact spectrum.
In the $1^0 2^2$ and $1^1 2^1$ spectra (second and fourth columns), the local and adiabatic harmonic models still perform quite well, with the LHA spectra capturing the small peaks at lower (i.e., more negative) wavenumbers slightly better.
 
 In general, the vertical harmonic model performs worse than the adiabatic and local harmonic approximations, particularly in terms of peak positions. At the same level of initial excitation in a single mode, the global harmonic models perform better when mode 2 is excited than when mode 1 is excited (compare the second and third columns of Figs.~\ref{fig:2d} and \ref{fig:2da} for $1^0 2^2$ and $1^2 2^0$ spectra). This is consistent with the fact that mode 1 is more `anharmonic' than mode 2 based on our parametrisation ($\chi_1 > \chi_2$). 
  Because the initial states with higher vibrational excitations are more delocalised and feel the anharmonic shape of the potential more, the spectra from all three approximations become worse at higher vibrational excitations (for example, compare the $1^0 2^0$ and $1^2 2^0$ spectra or the $1^1 2^1$ and $1^3 2^1$ spectra).

The spectra from the adiabatic model capture the high-frequency (i.e., less negative) peaks in the more harmonic regions more accurately, whereas the local harmonic dynamics reproduces better the peaks in the more anharmonic low-frequency tail region. For example, in the $1^{3}2^{1}$ spectra, the adiabatic model agrees well with the quantum result for the peaks in the region $>-5000\,\text{cm}^{-1}$ but the LHA is better at lower wavenumbers. We also observe several small, unphysical negative peaks in the LHA spectra owing to the nonconservation of energy and nonlinearity of the TDSE in the local harmonic dynamics \cite{Wehrle_Vanicek:2015,Begusic_Vanicek:2019,Vanicek:2023}. Unphysical peaks at positive $\Delta \tilde{\nu}$ also appear and will be discussed in more detail  in 
Sections \ref{subsec:anthracene} and \ref{subsec:cf2}.

Nonetheless, compared to the results of the global harmonic models, the local harmonic results differ less from the exact spectra, especially in the lower frequency region.  This region corresponds to transitions to higher vibrational levels, which are more significantly affected by the anharmonicity of the ground-state PES. In contrast, the adiabatic harmonic model tends to perform slightly better in the higher-frequency range for transitions to vibrational levels closer to the harmonic region. 

This numerical example made it possible to compare the local harmonic Hagedorn spectra with exact quantum benchmarks, without additional complications from electronic structure calculations or experiments. Yet, the main advantage of the local harmonic method is that it can be efficiently combined with on-the-fly ab initio dynamics in real molecules. In the following, we therefore present on-the-fly ab initio local harmonic Hagedorn wavepacket calculations of the SVL spectra of anthracene and difluorocarbene.

\subsection{Anthracene}
\label{subsec:anthracene}

In the first implementation of a time-dependent approach to SVL spectra, Tapavicza used a global harmonic model of anthracene to demonstrate a generating-function-based method for treating spectra arising from singly excited vibrational levels \cite{Tapavicza:2019}.
In our earlier work \cite{Zhang_Vanicek:2025a}, we applied \emph{global harmonic} Hagedorn wavepacket dynamics to evaluate the $\mathrm{^1B_{2u}}\to \mathrm{^1A_g}$ SVL spectra of anthracene beyond singly excited levels. Good agreement with the experimental results was achieved, with some discrepancies.
In a perfectly harmonic system, the adiabatic and vertical harmonic approximations should yield identical spectra. However, despite good agreement of the AHA with the anthracene experiment, noticeable differences were observed between the AHA and VHA results. This raised the question whether anharmonicity played a significant role in the observed deviations. Here, we therefore carry out the local harmonic Hagedorn dynamics to assess the potential anharmonicity contributions.

The efficiency of the local harmonic Hagedorn approach makes it possible to partially include anharmonicity by performing a single on-the-fly ab initio calculation for a system with 66 vibrational degrees of freedom.
The initial wavepacket was constructed from the optimized geometry and Hessian of the $\mathrm{^1B_{2u}}$ excited state, calculated with Gaussian 16 \cite{software_g16} using linear-response time-dependent density functional theory with the PBE0 functional \cite{Adamo_Barone:1999} and the def2\nobreakdash-TZVP basis set \cite{Weigend_Ahlrichs:2005} (see the supporting data \cite{data_Zhang_Vanicek:2025a} of Ref. \cite{Zhang_Vanicek:2025a}).
The parameters of the Gaussian wavepacket were propagated with a time step of 8 au for a total time of $8 \times 10^4$ au $(\sim 1.9\,\text{ps})$ using a second-order TVT geometric integrator \cite{book_Lubich:2008,Vanicek:2023}. At each point along the classical trajectory of the wavepacket's centre, the potential energy, its gradient, and its Hessian were computed with Gaussian 16 \cite{software_g16} in Cartesian coordinates at PBE0/def2\nobreakdash-TZVP level of theory. These quantities were then transformed into normal-mode coordinates to construct the local harmonic potential employed for the propagation. The autocorrelation functions of the Hagedorn wavepackets were computed every four steps using the algebraic
algorithm described in Ref.~\cite{Vanicek_Zhang:2025}. A Gaussian damping function with a half-width at half-maximum
of 20000 au was applied to the autocorrelation functions before performing the
Fourier transform.

\begin{figure}
    \centering
    \includegraphics[width=0.9\linewidth]{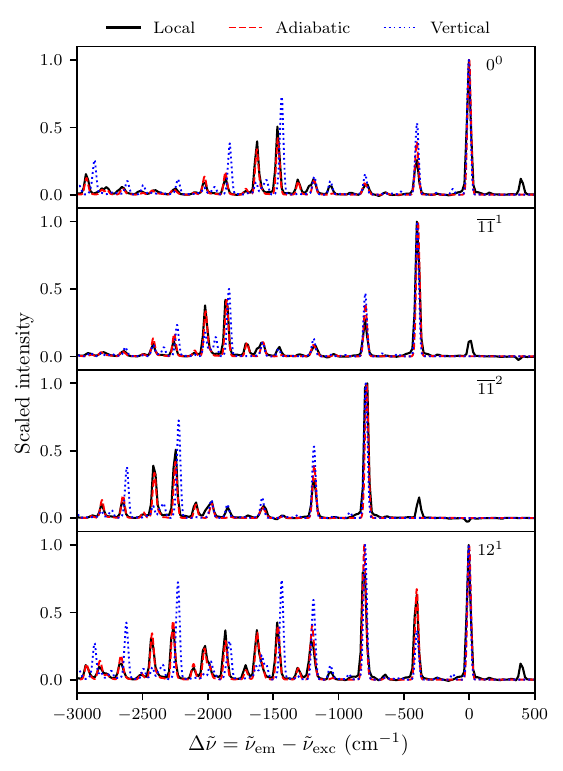}
    \caption{Comparison of SVL spectra computed using local (black solid lines), adiabatic (red dashed lines) and vertical (blue dotted lines) harmonic approximations from levels $0^0$, $\overline{11}^1$, $\overline{11}^2$, and $12^1$ of anthracene. The wavenumbers are not empirically scaled.}
    \label{fig:anth}
\end{figure}

We consider the emission spectra from the vibrational ground level ($0^0$) and 
 $\overline{11}^1$, $\overline{11}^2$, $12^1$ levels (see also Figure~4 in Ref.~\cite{Zhang_Vanicek:2025a}, where the $12^2$ spectrum was presented to validate the use of adiabatic harmonic approximation). The labels of the vibrational modes follow the convention in the experimental reference \cite{Lambert_Zewail:1984}; the bar in $\overline{11}$ indicates a mode of $\mathrm{b_{1g}}$ symmetry, whereas mode $12$ (without bar) is of $\mathrm{a_g}$ symmetry.
Comparison of the local, adiabatic, and vertical harmonic approximations in Figure~\ref{fig:anth} shows that the local harmonic SVL spectra of anthracene closely resemble the spectra computed using the adiabatic harmonic approximation but differ significantly from the the vertical harmonic spectra.
In the local harmonic spectra, very weak unphysical negative intensities appear for $\overline{11}^1$ and $\overline{11}^2$ levels (due to the nonlinearity of the effective Hamiltonian), while spurious peaks around $400\,\text{cm}^{-1}$ are visible for the $0^0$, $\overline{11}^1$, and $12^1$ levels.
These unphysical peaks at positive $\Delta\tilde{\nu}$ (i.e., apparent emission at higher energy than the initial excitation) appear because the state-dependent local potentials along the LHA trajectory may sometimes have vibrational levels with lower energy than the true potential's zero-point energy.

In the case of anthracene, accounting for Hessian changes along the classical trajectory in the local harmonic dynamics does not produce significantly different spectra compared to the adiabatic harmonic results, indicating that anharmonicity plays only a limited role as long as one chooses the `correct' global harmonic model. In anthracene, the adiabatic harmonic approximation provides a better description than the vertical harmonic one with the chosen density functional and basis set (PBE0/def2-TZVP). This is true quite generally, but not always. For example, in the case of ammonia, it is the vertical harmonic approximation that yields a more correct absorption spectrum \cite{Wehrle_Vanicek:2015,Kletnieks_Vanicek:2023}. Therefore, if discrepancy appears between the adiabatic and vertical harmonic spectra in a system believed to be harmonic, the local harmonic Hagedorn wavepacket dynamics may help to make the final verdict. 

\subsection{Difluorocarbene}
\label{subsec:cf2}

Whereas the partial inclusion of anharmonicity in anthracene with local harmonic dynamics did not lead to significant changes from the adiabatic global harmonic model, the situation is different for more flexible molecules with more anharmonic potential energy surfaces.
In Ref.~\cite{Zhang_Vanicek:2025}, we used ab initio local harmonic Hagedorn wavepacket dynamics to compute the SVL fluorescence spectra of difluorocarbene (CF$_2$) arising from excited bending-mode (mode 2) levels (from $2^0$ to $2^6$) and obtained an excellent agreement with experiment, particularly at lower excitation levels. In contrast, both vertical and adiabatic harmonic spectra were much worse not only for the doubly excited $2^2$ level but even for the fluorescence spectrum from the vibrational ground $2^0$ level (see Figure~1 in Ref.~\cite{Zhang_Vanicek:2025}). 

\begin{figure*}
    \centering
    \includegraphics[width=\linewidth]{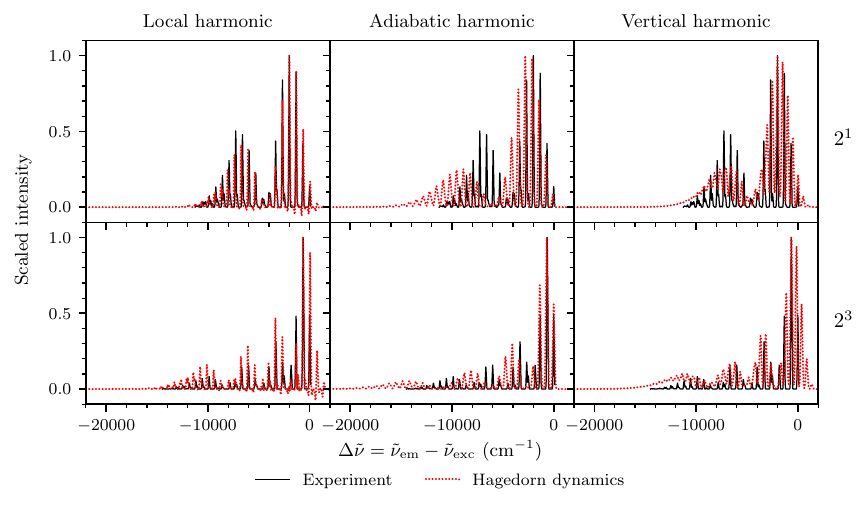}
    \caption{Comparison of the experimental (black solid line) SVL fluorescence
spectra from $2^1$ and $2^3$ levels of CF$_2$ from Ref.~\cite{King_Stephenson:1979} with computed (red dotted line) spectra
using the local, adiabatic, and vertical harmonic Hagedorn wavepacket
dynamics.}
    \label{fig:cf2}
\end{figure*}

Here, we performed a similar analysis for the
$2^1$ and $2^3$ spectra. The adiabatic, vertical, and local harmonic spectra from levels $2^1$ and $2^3$ are compared to the experimental spectra in Figure~\ref{fig:cf2} and were obtained using the same computational approach and wavenumber shifting procedure as described in Ref.~\cite{Zhang_Vanicek:2025} and its supplemental material.
The vertical harmonic approximation describes the envelope and its splitting reasonably well; however, similar to the local harmonic spectra, it produces unphysical peaks at positive $\Delta\tilde{\nu}$.
These occur because the harmonic expansion at the Franck--Condon point in CF$_2$ results in vibrational levels at energies lower than the zero-point energy of the true potential.
Nonetheless, the global harmonic spectra still display, at least qualitatively, the signature splitting of the spectral envelope due to initial vibrational excitation, and we now explain it analytically using a reduced-dimensional model.

The splitting of the spectral envelope, which does not appear in standard (i.e., ground-level) fluorescence, arises from motion in the bending mode. CF$_2$ belongs to the
$C_{2v}$ point group and has three fundamental vibrational modes:
symmetric stretching ($ \mathrm{a_1}$), bending  ($\mathrm{a_1}$), and asymmetric stretching ($\mathrm{b_2}$). 
Within the harmonic approximation, the asymmetric stretching mode shows no displacement between the electronic states and is thus inactive in the
$\tilde{A}\mathrm{^1B_1} \to \tilde{X}\mathrm{^1A_1}$ transition within the Condon approximation.
Of the two allowed modes, the bending mode has a much larger displacement than 
the symmetric stretching one, as the change in bond angle between the 
optimised geometries of the two electronic states is greater than the change in 
bond lengths. The bending progression thus dominates the spectra.

\begin{figure}
    \centering
    \includegraphics[width=0.9\linewidth]{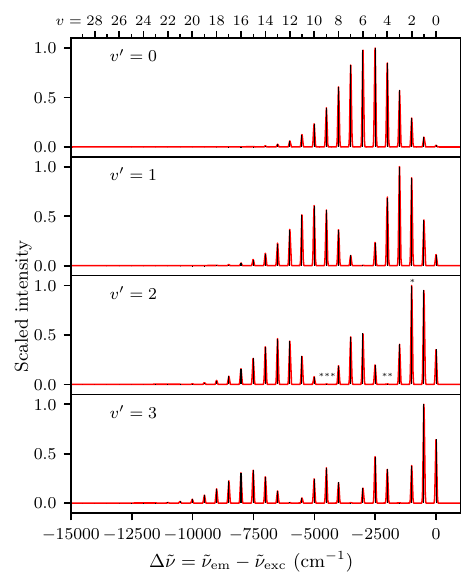}
    \caption{Comparison between SVL spectral intensities computed with the analytical formula [Equation~(\ref{eq:1d_analytical}), black sticks] and with the Hagedorn approach (red solid lines) in a reduced one-dimensional displaced harmonic oscillator model based on CF$_2$ for emission from excited bending-mode levels with initial vibrational quantum numbers $v^\prime = 0, 1, 2, 3$; the final vibrational levels $v$ are indicated along the top axis. Transitions marked by *, **, and *** are analyzed in Figure \ref{fig:cf2_fc}.}
    \label{fig:cf2_1d}
\end{figure}

Figure~\ref{fig:cf2_1d} shows the SVL spectra computed using a reduced one-dimensional displaced harmonic oscillator model with a Huang--Rhys factor $S=m \omega \Delta q^2 / 2 \hbar = 6.15$. The displacement $\Delta q$ in the bending mode was obtained from the adiabatic harmonic model constructed from PBE0/aug-cc-pVTZ calculations (see the Supplemental Material to Ref.~\cite{Zhang_Vanicek:2025a}), and we assume that the bending mode has the same frequency $500.84\,\mathrm{cm}^{-1}$ in the ground electronic state as in the excited state (i.e., $\omega_g = \omega_e$).
In a one-dimensional displaced harmonic oscillator system, the intensity $\mathcal{P}_{v\leftarrow v'}$ of vibronic transition from an excited vibrational level $v'$ in the excited electronic state to the vibrational level $v$ in the ground electronic state can be evaluated analytically as \cite{Wagner:1959}
\begin{equation}
   \mathcal{P}_{v\leftarrow v'} = e^{-S}\,
\frac{\min(v,v')!}{\max(v,v')!}\,
S^{|v'-v|}\,
\left[ L_{\min(v,v')}^{(|v'-v|)}(S) \right]^2.
\label{eq:1d_analytical}
\end{equation}
In particular, $\mathcal{P}_{v\leftarrow v'}$ follows a Poisson distribution modulated by associated Laguerre 
polynomials $L^{(j)}_k(x)=\sum_{m=0}^{k} (-1)^m \binom{k+j}{k-m} \frac{x^m}{m!}$,
which give rise to the observed splittings.

\begin{figure}
    \centering
    \includegraphics[width=0.9\linewidth]{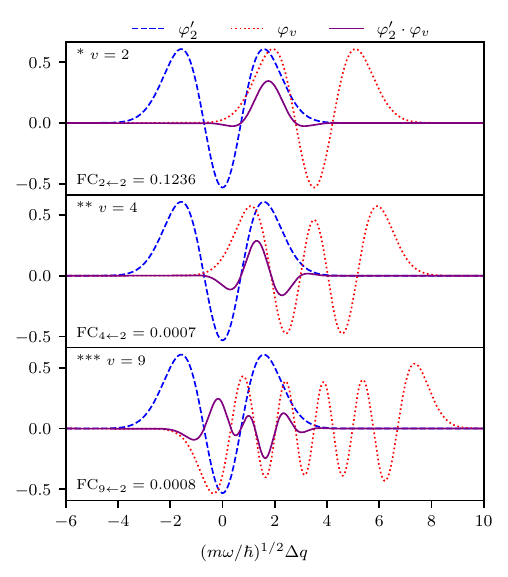}
    \caption{Initial ($\varphi^\prime_{2}$, red dotted lines) and final ($\varphi_{v}$, blue dashed lines) vibrational wavefunctions, as well as their products (purple solid lines), as a function of the dimensionless displacement $(m \omega / \hbar)^{1 / 2} \Delta q$, in the transitions from $v^\prime = 2$ to $v=2,4,9$ (indicated, respectively, by *, **, *** in Figure \ref{fig:cf2_1d}); the integrated Franck--Condon factors $\text{FC}_{v\leftarrow2} = \left|\langle \varphi^\prime_{2} \mid \varphi_v \rangle\right|^2$ are shown at the bottom left of each panel.}
    \label{fig:cf2_fc}
\end{figure}

The splitting of the envelope can also be rationalized from a time-independent perspective.  As an example, Figure \ref{fig:cf2_fc} shows the vibrational eigenfunctions and the Franck--Condon integrands in the 2$ \leftarrow$2 (peak * in the third row of Figure \ref{fig:cf2_1d}), 4$\leftarrow$2 (peak **), and 9$ \leftarrow$2 (peak ***)  transitions. When the initial vibrational level is excited ($v^\prime > 0$), its wavefunction has $v^\prime$ nodes where its amplitude changes sign.
When the initial and final wavefunctions are in phase in the overlap region, as in the 2$\leftarrow$2 transition (first row of Figure \ref{fig:cf2_fc}), the Franck--Condon factor is large and the peak intensity is maximized.
 At a splitting point, the initial and final wavefunctions become unfavourably aligned, causing their product to alternate between positive and negative regions with roughly equal but opposite contributions that cancel out in the Franck--Condon integral, resulting in minimal spectral intensity, as seen in the cases of the 4$\leftarrow$2 (second row) and 9$\leftarrow$2 (third row) transitions.
 
It is comforting that the time-dependent Hagedorn approach gives exactly the same relative spectral intensities as those predicted by Equation~(\ref{eq:1d_analytical}) (compare the red lines and black sticks in Figure~\ref{fig:cf2_1d}). This agreement not only validates our Hagedorn wavepacket approach, but also confirms the correctness of the analytical equation~(\ref{eq:1d_analytical}).
While this reduced model for the displaced bending mode explains the splitting induced by vibrational excitation, the local harmonic approach provides a robust way to include anharmonic contributions that are required for a quantitative agreement with experiments.

\section{Conclusions}

We have described an efficient approximate approach for including anharmonicity effects in calculations of SVL spectra. In a two-dimensional coupled Morse potential, 
we validated this so-called `local harmonic Hagedorn wavepacket dynamics' against numerically exact quantum calculations free of electronic structure (and experimental) errors.
With relatively low initial vibrational excitations
in one or multiple modes, local harmonic dynamics effectively captures anharmonic effects in the SVL spectra of moderately anharmonic
and even nonseparable
systems. Although the differences between the exact and local harmonic spectra become more pronounced for higher initial excitations (since the delocalisation of excited vibrational wavepackets increases the effect of anharmonicity on the spectra), the LHA offers notable improvements over vertical and adiabatic harmonic approximations. 

In the examples presented, we assumed a harmonic initial potential energy surface and focused only on capturing the anharmonic effects in the final electronic state, which are generally more significant since the Franck--Condon point can be far from equilibrium.
However, this assumption becomes less valid at higher vibrational excitations or in molecules with strongly anharmonic excited-state surfaces.
Since Hagedorn functions form an orthonormal basis \cite{Hagedorn:1998,book_Lubich:2008}, Hagedorn wavepackets (linear combinations of Hagedorn functions) can, in principle, represent any vibrational state. Incorporating anharmonicity of the initial surface by projecting an anharmonic eigenstate (obtained by other techniques \cite{Kosloff_Tal-Ezer:1986,Meyer_Gatti:2006,Ceotto_Valleau:2011,Micciarelli_Ceotto:2018,Felker_Bacic:2019,Felker_Bacic:2020}) onto the Hagedorn basis is a subject of ongoing work.

The Hagedorn wavepacket dynamics has the advantage of obtaining SVL spectra from all initial vibrational levels using a single common Gaussian wavepacket trajectory, thus avoiding the need to repeat expensive electronic structure calculations for different initial excitations. Because this trajectory requires only local information of the potential, the method is well suited for on-the-fly ab initio implementations, even in larger molecules such as anthracene.

The importance of anharmonicity differs in different molecular systems, as illustrated by the examples of anthracene and difluorocarbene.
The local harmonic approach can be used to evaluate the adequacy of harmonic models and to indicate when additional anharmonic contributions may need to be considered, even when exact spectra are not accessible.

Hagedorn wavepackets describe nuclear vibrational dynamics in normal-mode coordinates. It is most suitable for simulating vibrationally resolved electronic spectra. In ab initio applications and on-the-fly implementation, rovibrational couplings are neglected during the coordinate transformations.
For highly accurate calculations that resolve rovibrational levels observable in SVL or other spectroscopies, more advanced quantum dynamical approaches, such as those developed by Ba\v{c}i\'c and co-workers \cite{Bacic_Light:1989,Mladenovic_Bacic:1990,Zhang_Bacic:1995, Felker_Bacic:2019,Simko_Bacic:2025}, are required, particularly for highly anharmonic, flexible systems or when higher (ro)vibrational excitations are of interest.

In on-the-fly local harmonic dynamics, the computation of Hessians is the most computationally demanding part of the calculation and can become challenging in out-of-equilibrium regions.
To further increase the efficiency of the local harmonic Hagedorn approach, techniques like Hessian interpolation \cite{Wehrle_Vanicek:2014} or the single Hessian approximation \cite{Begusic_Vanicek:2019} can be employed, similar to their use in simulating conventional absorption and emission spectra from ground vibrational levels with the thawed Gaussian approximation. Although we focused on single vibronic level spectroscopy, the local harmonic Hagedorn approach described here should be equally useful for evaluating other spectra requiring the propagation of non-Gaussian wavepackets, such as Herzberg--Teller spectra or hot-band contributions to spectra at nonzero temperatures.

\section*{Disclosure statement}

The authors report there are no competing interests to declare.

\section*{Funding}

The authors acknowledge financial support from the EPFL.

\bibliographystyle{aipnum4-2}
\bibliography{svl_lha_model_v17}

\end{document}